\documentclass[%
 reprint,
 amsmath,amssymb,
 aps, physrev,
 superscriptaddress,
 longbibliography
]{revtex4-2}

\usepackage{graphicx}
\usepackage{dcolumn}
\usepackage{bm}
\usepackage{xcolor}
\usepackage{wrapfig}
\usepackage{booktabs}

\begin{document}

\preprint{}

\title{Probing Spin Configurations in Exchange-Coupled Magnetic Bilayers with Orthogonal Anisotropies via Anomalous Hall and Nernst Effects}

\author{Sebin Jung}
\affiliation{Department of Applied Physics, University of Tsukuba, Tsukuba, Ibaraki 305-8573, Japan}

\author{Hiroki Koizumi} 
\affiliation{Department of Applied Physics, University of Tsukuba, Tsukuba, Ibaraki 305-8573, Japan}

\author{Hiroaki Sukegawa}
\affiliation{National Institute for Materials Science (NIMS), Tsukuba, Ibaraki 305-0047, Japan}
\author{Hideto Yanagihara}
\email{yanagihara.hideto.fm@u.tsukuba.ac.jp}
\affiliation{Department of Applied Physics, University of Tsukuba, Tsukuba, Ibaraki 305-8573, Japan}
\affiliation{Tsukuba Research Center for Energy Materials Science (TREMS), University of Tsukuba, Tsukuba, Ibaraki 305-8573, Japan}
\affiliation{Tsukuba Research Center for Organic-Inorganic Quantum Spin Science and Technology (OIQSST), University of Tsukuba, Tsukuba, Ibaraki 305-8573, Japan}
\date{\today}

\begin{abstract}
When two ferromagnetic thin films with different magnetic easy axes are coupled via the exchange interaction, the magnetization process becomes nontrivial. In this paper, we demonstrate that three-dimensional magnetization information in such a bilayer system can be accessed electrically by combining measurements of the anomalous Hall effect (AHE) and anomalous Nernst effect (ANE). Specifically, we investigate a CoFe$_2$O$_4$(001)/Fe bilayer, where the insulating nature of CoFe$_2$O$_4$ ensures that these transport measurements selectively probe only the conductive Fe layer. By combining the AHE and ANE results and comparing them with a simple micromagnetic simulation, we probe the magnetic configuration of antiferromagnetically coupled Fe layer and suggest the emergence of a twisted magnetic structure near the interface.
\end{abstract}

\maketitle

\section{Introduction}
Magnetic heterostructures have attracted significant interest both theoretically \cite{1970_Brodkorb, 1987_Yafet, 1991_Bruno, 1993_Stiles, 1997_Koon, 1999_Stiles} and experimentally \cite{1986_Mahkrzak, 1986_Salamon, 1996_Moran, 1999_Nogues, 2012_Bailey, 2024_Surampalli}. These systems are characterized by the interlayer exchange coupling (IEC) of multiple magnetic layers through mechanisms such as coupling induced by the variation in the electron density of states arising from the spin-dependent quantum wall or RKKY interaction \cite{1995_Bruno, 1956_Kasuya, 1991_Bruno, 1992_Johnson, 2012_Bailey}. Because these couplings affect the static magnetization configuration of a system \cite{1990_Parkin, 2020_olmos}, controlling their strength and direction enables field-free magnetization manipulation and offers a promising platform for spintronic devices. Such coupling phenomena have been widely harnessed in the development of giant magnetoresistance (GMR) devices \cite{1989_Grunberg, 2023_Liu} and synthetic antiferromagnet-based tunnel magnetoresistance (TMR) devices \cite{2004_Stobiecki, 2014_Yoshida}.

Despite the extensive research on magnetic heterostructures, most previous studies have focused on systems in which the magnetic preferential axes of the layers are parallel. In such configurations, the effect of IEC on the magnetic structure is well understood\cite{1990_Dieny}. However, the magnetic configuration of heterostructures in which the easy axes of the two ferromagnets are orthogonal (e.g., in-plane vs. out-of-plane) remains relatively unexplored. Nevertheless, from a recent application perspective, systems featuring orthogonal magnetizations have attracted significant interest for magnetic devices. For example, achieving a perpendicular magnetization alignment between adjacent ferromagnetic layers offers a highly effective approach for realizing magnetic sensors that operate in a wide magnetic field range\cite{2015_Lee}. Hence, observing and controlling the precise magnetization configurations that result from the balance between these orthogonally aligned uniaxial anisotropies and IEC is crucial to both fundamental \cite{2011_Nguyen, 2014_Nguyen, 2020_Zachary} and practical investigations \cite{2023_Lisik, 2024_Nakatani}.

In this study, we investigated the effect of IEC in a bilayer system composed of CoFe$_2$O$_4$ with strong perpendicular magnetic anisotropy (PMA)\cite{1995_BRABERS, 2013_Niizeki, 2017_Coey} and metallic Fe with soft in-plane anisotropy. Notably, the electrical transport measurements such as the anomalous Hall effect (AHE) and anomalous Nernst effect (ANE) selectively probe the magnetization of the Fe layer exclusively because the CoFe$_2$O$_4$ layer is electrically insulating. By consecutively measuring the AHE and ANE, the three dimensional magnetization of Fe moment can be determined from the measured signal of each direction. Herein, we report the direct measurement of the Fe magnetic moment configuration in response to the intricate balance of anisotropies, IEC, and external fields to in-plane and out-of-plane directions, respectively. Furthermore, a layered-spin model simulation suggested the existence of non-collinear spin texture within the bilayer interface.

\section{Experiment}

\begin{figure}[htb]
    \centering
    \includegraphics[width=1.0\columnwidth]{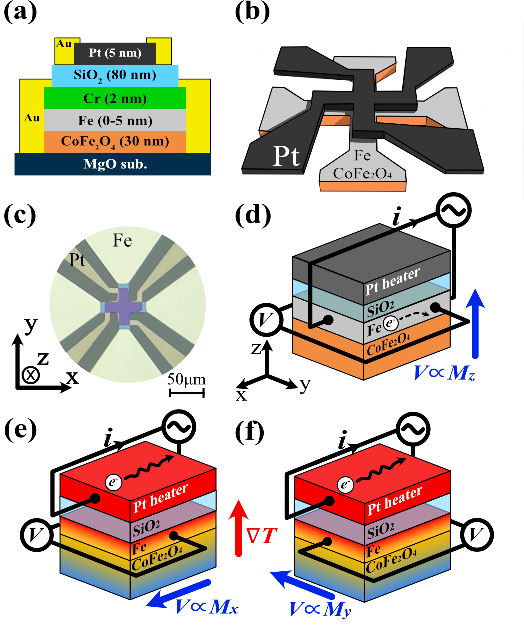} 
    \caption{Measurement setup and transport responses of a magnetic bilayer with orthogonal magnetic anisotropies. (a) Schematic illustration of the multi-layered sample. (b) Schematic illustration of the Hall bar device pattern with an integrated Pt heater. (c) Optical photograph of the patterned sample. (d) Configurations for AHE measurement. (e, f) Configurations for ANE measurements, where the signals are proportional to the (e) $M_x$ and (f) $M_y$ components, respectively. By simply changing the current paths and voltage contacts, three-dimensional magnetization of a conductive magnetic layer can be probed.}
    \label{fig1}
\end{figure}

A sample stack of CoFe$_2$O$_4$(30 nm)/Fe (0-5 nm)/Cr(2 nm)/SiO$_2$(80 nm)/Pt(5 nm) was grown via a (reactive) sputtering on a MgO substrate (Fig. \ref{fig1}(a)) and subsequently patterned into a Hall bar structure with an integrated Pt heater (Fig. \ref{fig1}(b, c)). The details of the growth processes of CoFe$_2$O$_4$ are given in Ref.~[\onlinecite{2013_Niizeki}]. To probe the detailed magnetization configurations, we conducted AHE and ANE measurements consecutively by simply switching the current path and the amplitude (Figs. \ref{fig1}(d-f)). The device structure ensures that the transport signals originate exclusively from the conductive Fe layer owing to the electrically insulating nature of CoFe$_2$O$_4$. The AHE signal, measured by applying an in-plane current to the Fe layer, detects the out-of-plane magnetization component ($M_z$). In contrast, the ANE signal, induced by a perpendicular temperature gradient to the film generated by the Joule heat from the Pt heater, detects the in-plane magnetization components ($M_x$ and $M_y$). By normalizing each signal to its corresponding saturation value under high magnetic fields, we determined the magnetization configurations for both the out-of-plane and in-plane directions within the Fe layer.

\section{Results and discussion}
\begin{figure}[htbp]
    \centering
    \includegraphics[width=1.0\columnwidth]{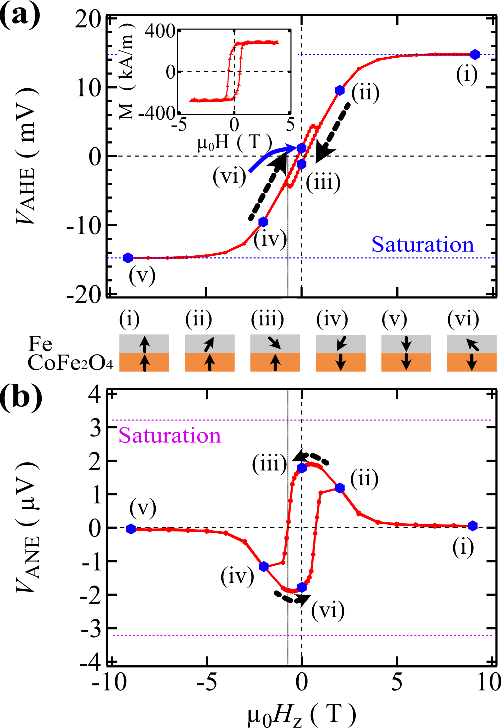} 
    \caption{Transport responses of the magnetic bilayer with orthogonal magnetic anisotropies. (a) AHE voltage and (b) ANE voltage (Fig. \ref{fig1} (e)) measured under an out-of-plane magnetic field ($H_z$). The middle panels depict the spin configurations corresponding to the magnetization reversal process (i–vi). The inset in (a) presents the VSM curve of a single CoFe$_2$O$_4$ layer on MgO under an out-of-plane direction magnetic field.}
    \label{fig2}
\end{figure}

For the sample with a 1.5-nm-thick Fe layer, Figs. \ref{fig2}(a) and (b) present the consecutive AHE and ANE measurements under a magnetic field applied along the out-of-plane direction, respectively. The ANE signals in Fig. \ref{fig2}(b) were obtained using the measurement configuration shown in Fig. \ref{fig1}(e). The blue and purple dashed lines represent the saturation levels of the AHE and ANE signals, corresponding to the magnetic moments fully aligned along the out-of-plane and in-plane directions, respectively. Under an out-of-plane magnetic field of $\mu_0H = $ 9.0 T, the AHE signal saturated whereas the ANE signal vanished, indicating the perpendicular saturation of the magnetic moment of the Fe layer (Fig. \ref{fig2}(a), point (i)). As the applied field decreased, the AHE signal decreased whereas the ANE signal increased. This behavior indicated the increase of the in-plane component of Fe magnetic moments toward the in-plane direction (Fig. \ref{fig2}(a), point (ii)). In the remanent state (Fig. \ref{fig2}(a), point(iii)), a negative AHE signal was observed, which corresponded to the inversion of the $M_z$ component of the Fe layer. This result suggested the existence of an antiferromagnetic IEC between the two magnetic layers. The exchange coupling constant could be determined by equating the Zeeman energy to the interface exchange energy between the remanent state and the coercive field, assuming the CoFe$_2$O$_4$ magnetization remained fixed along the perpendicular direction (for details, see Ref. [\onlinecite{2020_Koizumi}]), and it was estimated to be $-0.5$ mJ/m$^2$. Furthermore, as the magnetic field increased in the $-z$ direction, the sign of the ANE signal reversed at the coercivity of the CoFe$_2$O$_4$ layer(see the inset of Fig. \ref{fig2}(a)), $\mu_0H = $ 0.6 T, indicating that the canting direction of the Fe layer was related to the IEC with CoFe$_2$O$_4$ (Fig. \ref{fig2}(a), point (iv)). The ANE signals obtained using the measurement configuration shown in Fig. \ref{fig1}(f) are presented in Sec. II of the supplementary material.

\begin{figure}[tbp]
    \centering
    \includegraphics[width=1.0\columnwidth]{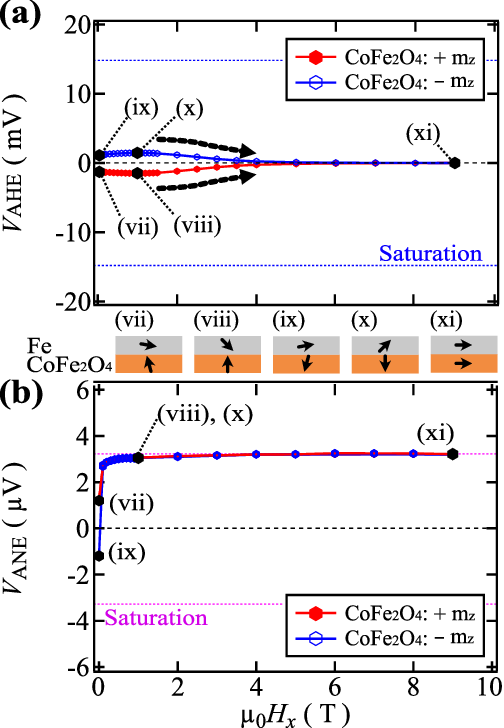} 
    \caption{Transport responses of the magnetic bilayer with orthogonal magnetic anisotropies. (a) AHE voltage and (b) ANE voltage (Fig. \ref{fig1}(e)) measured under an in-plane magnetic field ($H_x$). The middle panels illustrate the spin configurations corresponding to the magnetization reversal process (vii-xi).}
    \label{fig3}
\end{figure}

For the same device, Figs. \ref{fig3}(a) and (b) present the AHE and ANE signals consecutively measured under a magnetic field applied along the $+x$ direction. The ANE signals in Fig. \ref{fig3}(b) were obtained using the measurement configuration shown in Fig. \ref{fig1}(e). Before the measurement, the CoFe$_2$O$_4$ layer was saturated along the $+z$ and $-z$ directions (Fig. \ref{fig2}, point (i) or (v)) to align the magnetization in one direction as an initial state, after which the out-of-plane direction magnetic field was removed (Fig. \ref{fig2}, point (iii) or (vi), respectively). In this state, the sign of the AHE signal depended on the remanent magnetization direction of the CoFe$_2$O$_4$ (Fig. \ref{fig3}, points (vii) and (ix)). The results suggested that the slight canting of the Fe moments was induced by the antiferromagnetic IEC. The ANE signal, however, exhibited the same behavior regardless of the initial saturation direction of CoFe$_2$O$_4$, indicating the same $M_x$ of the canted Fe moments for a given in-plane field. Notably, the magnitude of the AHE signal increased up to an in-plane field of $\mu_0H = $ 1.0 T (Fig. \ref{fig3}(a), points (viii) and (x)), after which it decreased monotonically to zero at strong in-plane fields (Fig. \ref{fig3}(b), point (xi)), exhibiting a hump-shaped signal. This behavior indicated that the $M_z$ component of the Fe moment increased with the in-plane field up to $\mu_0H = $ 1.0 T, after which it saturated under a strong in-plane field.

\begin{figure}[tbp]
    \centering
    \includegraphics[width=1.0\columnwidth]{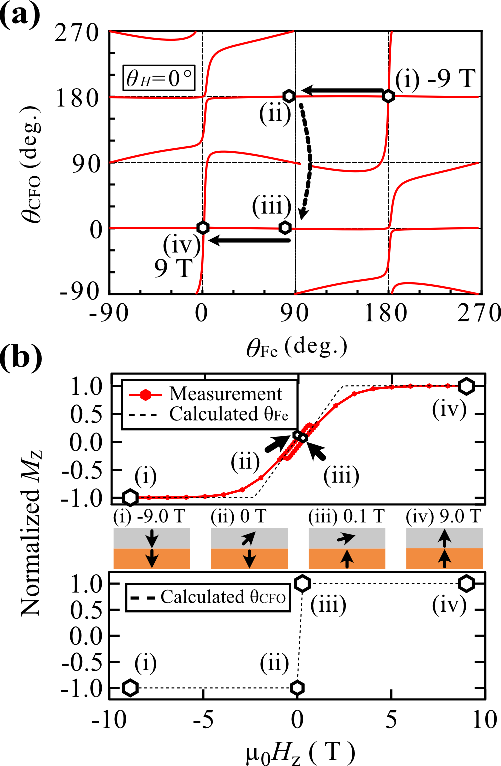} 
    \caption{Macrospin model analysis of the magnetization configuration in the CoFe$_2$O$_4$/Fe bilayer. (a) Relationship between $\theta_{Fe}$ and $\theta_{CFO}$ satisfying Eq.~\ref{equ2} under an out-of-plane magnetic field $(\theta_{H} = 0)$. (b) Normalized out-of-plane magnetization ($M_z$) derived from AHE measurements, and the $M_z$ components of each layer estimated using the macrospin model assumption under an out-of-plane magnetic field. Points (i)–(iv) indicate specific magnetic states during the out-of-plane field sweep from $-9$~T to $9$~T.}
    \label{fig4}
\end{figure}

Based on the physical parameters extracted from the combined AHE and ANE measurements, the magnetization process of the coupled CoFe$_2$O$_4$/Fe bilayer can be reconstructed within a macrospin framework. To understand the field-dependent magnetization configurations of the two layers, the equilibrium magnetization directions were determined by minimizing the total free energy, equivalently by solving the zero-torque condition. The model consists of two ferromagnetic layers with orthogonal magnetic easy axes that are antiferromagnetically coupled at the interface. Furthermore, the magnetization vectors of both layers were assumed to lie in the $x$–$z$ plane, which is a reasonable approximation in the high-field regime. The free energy per unit area of the magnetic bilayer $\mathrm{E_{tot}}$ is expressed as,

\begin{equation}
\begin{split}
E_{\mathrm{tot}} &= -\mu_{0}H m_{\mathrm{Fe}}t_{\mathrm{Fe}}\cos(\theta_{H} - \theta_{\mathrm{Fe}}) + \frac{1}{2}\mu_{0}m_{\mathrm{Fe}}^{2}t_{\mathrm{Fe}}\cos^{2}\theta_{\mathrm{Fe}} \\- &\mu_{0}H m_{\mathrm{CFO}}t_{\mathrm{CFO}}\cos(\theta_{H} - \theta_{\mathrm{CFO}})- Kt_{\mathrm{CFO}}\cos^{2}\theta_{\mathrm{CFO}} \\&+\frac{1}{2}\mu_{0}m_{\mathrm{CFO}}^{2}t_{\mathrm{CFO}}\cos^{2}\theta_{\mathrm{CFO}}
- J\cos(\theta_{\mathrm{Fe}}-\theta_{\mathrm{CFO}}).
\end{split}
\label{equ1}
\end{equation}

Here, $\mu_0$ is the vacuum permeability; $m_X$ and $t_X$ are the magnetization and thickness of layer $X$ ($X$ = Fe or CoFe$_2$O$_4$), respectively; $\theta_{Fe}$, $\theta_{CFO}$, and $\theta_{H}$ are the polar angles of the Fe, CoFe$_2$O$_4$ magnetization, and magnetic field relative to the $z-$axis, respectively; $K$ is the magneto-crystalline anisotropy constant of CoFe$_2$O$_4$; and $J$ is the exchange coupling constant. 

By differentiating the total energy $\mathrm{E_{tot}}$, we can determine the torque exerted on the magnetic moments of each layer. This derivation establishes the relationship between $\theta_{\mathrm{Fe}}$ and $\theta_{\mathrm{CFO}}$ (Eq. (\ref{equ2})), and the identity to be satisfied when two magnetic layers are in equilibrium is,

\begin{align}
&J\sin(\theta_{\mathrm{Fe}}-\theta_{\mathrm{CFO}}) \{C\sin(\theta_{H}-\theta_{\mathrm{CFO}}) + A\sin(\theta_{H}-\theta_{\mathrm{Fe}})\} \nonumber \\
&= BC\sin\theta_{\mathrm{Fe}}\cos\theta_{\mathrm{Fe}}\sin(\theta_{H}-\theta_{\mathrm{CFO}}) \nonumber \\
&\quad - AD\sin(\theta_{H}-\theta_{\mathrm{Fe}})\sin\theta_{\mathrm{CFO}}\cos\theta_{\mathrm{CFO}}, \label{equ2} \\
\intertext{where}
A &= \mu_0 m_{\mathrm{Fe}} t_{\mathrm{Fe}}, \quad
B = \mu_0 m_{\mathrm{Fe}}^2 t_{\mathrm{Fe}}, \nonumber \\
C &= \mu_0 m_{\mathrm{CFO}} t_{\mathrm{CFO}}, \quad
D = \mu_0 m_{\mathrm{CFO}}^2 t_{\mathrm{CFO}} - 2K t_{\mathrm{CFO}}. \nonumber
\end{align}

\begin{figure}[tbp]
    \centering
    \includegraphics[width=1.0\columnwidth]{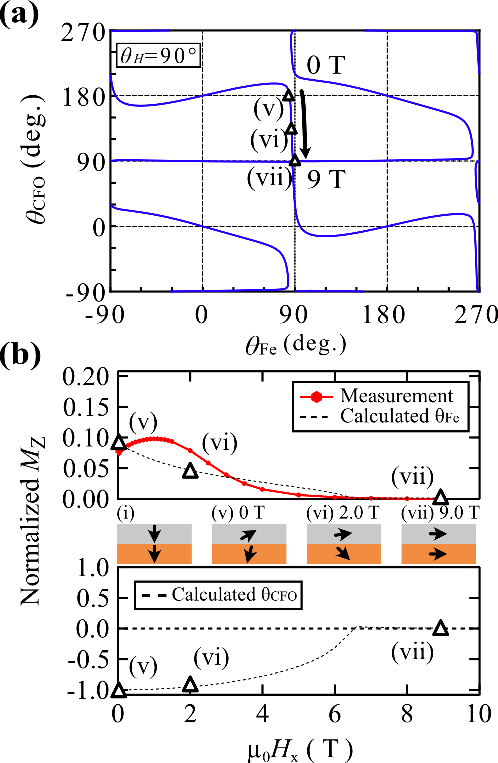} 
    \caption{Macrospin model calculation of the magnetization configuration in the CoFe$_2$O$_4$/Fe bilayer. (a) Relationship between $\theta_{Fe}$ and $\theta_{CFO}$ satisfying the equilibrium condition under an in-plane magnetic field ($\theta_{H}= \pi/2$). (b) Experimental $M_z$ values compared with the $M_z$ components estimated using a macrospin model. Points (v)-(vii) indicate specific magnetic states during the in-plane field sweep from $0$~T to $9$~T.}
    \label{fig5}
\end{figure}

Figures \ref{fig4}(a) and \ref{fig5}(a) show the points satisfying Eq. (\ref{equ2}) under out-of-plane and in-plane magnetic fields, respectively. To prevent the system from stabilizing in a metastable state, we performed these calculations at angles slightly off the exact symmetry axes. The calculations were performed using the following parameters: $m_{\mathrm{Fe}} = 1700$ kA/m, $m_{\mathrm{CFO}} = 300$ kA/m, $t_{\mathrm{Fe}} = 1.5$ nm, $t_{\mathrm{CFO}} = 30$ nm, $K_{\mathrm{CFO}} = 1$ MJ/m$^3$, and $J = -0.5$ mJ/m$^2$. Based on these relationships, the alignment of the magnetic moment of CoFe$_2$O$_4$ and Fe layers were determined by minimizing Eq. (\ref{equ1}). 

Figure \ref{fig4}(b) presents the normalized $M_z$ values of the Fe layer derived from AHE measurements, and the $M_z$ component of each layer estimated using the macrospin model under an out-of-plane magnetic field. When a large magnetic field was applied to the system along the $-z$ direction, the magnetic moments of both layers saturated along the external field (Fig. \ref{fig4}, point (i)). As the magnetic field was reduced, the macrospin simulation estimated that the magnetic moment of the Fe layer canted toward the in-plane direction, whereas that of the CoFe$_2$O$_4$ layer remained aligned with the $-z$ axis (Fig. \ref{fig4}, point (ii)). As the magnetic field was increased in the $+z$ direction, the magnetic moment of the CoFe$_2$O$_4$ layer reversed. During this process, the macrospin model successfully captured the system overcoming the energy barrier and the subsequent tilting of the Fe layer's magnetization (Fig. \ref{fig4}, point (iii)). A strong magnetic field completely saturated the magnetic moments of the bilayer (Fig. \ref{fig4}, point (iv)). 

Figure \ref{fig5}(b) presents the $M_z$ values of the Fe layer obtained from the measurement and the $M_z$ values of the Fe layer estimated using the macrospin model under an in-plane magnetic field. The initial remanent state was set with the angles $\theta_{Fe}$ and $\theta_{CFO}$ at 90 and 180 degrees, respectively. In the remanent state, the macrospin model estimated that the magnetic moment of the Fe layer was slightly canted toward the in-plane direction, which can be expressed by the point where the red and blue lines in Figs. \ref{fig4}(a) and \ref{fig5}(a) overlap (Fig. \ref{fig5}, point (v)). However, unlike the measurement, the macrospin simulation estimated a monotonic saturation of the magnetic moments of the Fe and CoFe$_2$O$_4$ layers in response to the in-plane magnetic field (Fig. \ref{fig5}, point (vi)). This discrepancy suggested that the macrospin model, which assumes a uniform magnetic moment across the single layer, can capture only a limited aspect of the magnetic system's behavior and that a non-uniform spin structure in the layers must be considered \cite{2015_Yanagihara}.

\begin{figure*}[htbp]
    \centering
    \includegraphics[width=1.0\textwidth]{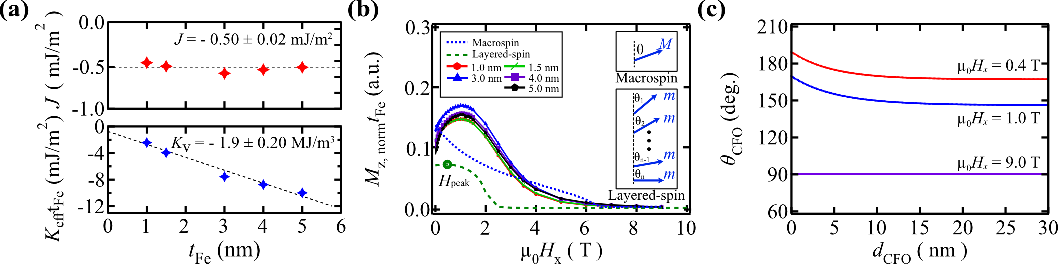} 
    \caption{Interfacial effects and layered-spin model simulation of the non-collinear spin texture in the CoFe$_2$O$_4$/Fe bilayer. (a) Exchange coupling constant and effective anisotropy energy as a function of Fe layer thickness. (b) The z-direction areal magnetization component as a function of the external magnetic field, estimated from AHE measurements for various Fe thicknesses. The green and blue dashed lines indicate the areal magnetization component simulated under the layered-spin and macrospin assumptions, respectively. The inset schematically depicts the macrospin model, which treats each magnetic layer as a single uniform magnetic moment, and the layered-spin model, where the magnetic moment of each atomic layer has the degree of freedom to rotate within the $x$-$z$ plane. (c) Simulated depth-dependent magnetic moment configurations across the bilayer. The x-axis represents the position of the layer relative to the CoFe$_2$O$_4$/Fe interface.}
    \label{fig6}
\end{figure*}

Fig. \ref{fig6}(a) presents the exchange coupling constant ($J$) of the magnetic bilayer and effective anisotropy energy ($K$) for the various Fe thicknesses from AHE measurements under an out-of-plane magnetic field. The exchange coupling constant of the system was determined as explained in Fig. \ref{fig2}. The effective anisotropy energy was determined by simply considering the shift of the exchange field. The results confirmed that the exchange coupling constant remained independent of the Fe thickness, whereas the shape anisotropy energy was estimated to be $-1.9$ MJ/m$^3$, which agreed closely with the shape anisotropy energy derived from its saturation magnetization $K_{\mathrm shape} = \frac{\mu_0}{2} m^2_{\mathrm Fe}$. Figure \ref{fig6}(b) presents the $M_z$ component of the areal magnetization as a function of the external in-plane magnetic field for various Fe thicknesses. The signals for all Fe thicknesses agreed closely, suggesting that the underlying mechanism can be attributed not to the bulky effect but to the interfacial interactions (see also Sec. III of the supplementary material).

As mentioned earlier, the macrospin model calculation successfully reproduced the overall magnetization processes of both Fe and CoFe$_2$O$_4$ layers. However, it could not reproduce the hump-like feature observed at finite magnetic fields, as shown by the blue dashed line in Fig. \ref{fig6}(b). To further investigate this anomalous behavior, we performed magnetic moment simulations under the layered-spin assumption (Inset of Fig. \ref{fig6} (b)), where the magnetic moments could rotate independently with distinct degrees of freedom per layer using the MuMax3 package (Ref.[\onlinecite{2014_Vansteenkiste_Mumax3}]). The green dashed line in Fig. \ref{fig6}(b) shows the simulated $M_z$ component of the areal magnetization of the Fe layer, under the layered-spin assumption. For the layered-spin assumption, the magnitude of the simulated ${M}_z$ component of the Fe layer exhibited a behavior similar to the hump-shaped feature in Fig. \ref{fig3}(a): an initial increase in $M_z$ up to $\mu_0H= $ 0.4 T followed by a decrease to 0 as the in-plane field increased. In contrast, for the macrospin simulation the results revealed a monotonic decrease in the ${M}_z$ component of Fe layer with increasing in-plane field. This discrepancy suggested that the non-monotonic behavior of the ${M}_z$ component under an in-plane field may be attributed to a non-trivial spin texture formed within the system.

Figure \ref{fig6}(c) shows the simulated magnetic moments across the CoFe$_2$O$_4$ using the layered-spin assumption. At an external field of $\mu_0H= $ 0.4 T, the simulation results revealed that a noncollinear spin texture was generated by the balance between the anisotropy, IEC and the Zeeman energy. Because the influence of the IEC is stronger near the interface, it generated a depth-dependent, inhomogeneous magnetic configuration. However, the calculation results showed only a slight magnetic twist within the Fe layer. This can be attributed to the fact that its thickness was thinner than the exchange coupling length of Fe. As the external magnetic field increased, the overall magnetization of the CoFe$_2$O$_4$ layer gradually saturated, yielding a uniform spin texture across the film. This behavior demonstrated that spin texture-controllable system can be achieved using the magnetic bilayers with orthogonal magnetic easy axes.

\section{Conclusion}
In this study, we investigated the detailed magnetization structure of a magnetic bilayer system with orthogonal magnetic anisotropy via combined AHE and ANE measurements. By employing continuous measurements of the AHE and ANE, we quantified the influence of exchange coupling on the magnetization and analyzed the non-collinear magnetization structure of the magnetic moments. Our findings suggest that this material system offers a promising platform for developing controllable twisted spin textures, where the chirality of the spin structure can be effectively manipulated via external magnetic fields.

\begin{acknowledgments}
This work was supported by the Japan Society for the Promotion of Science (JSPS) KAKENHI (22H04966, 23K26535, 24H00408, and 26K01016), and in part by the Advanced Research Infrastructure for Materials and Nanotechnology in Japan (ARIM) of the Ministry of Education, Culture, Sports, Science and Technology (MEXT), Japan (grant numbers JPMXP1224BA0008, JPMXP1225BA0030, and JPMXP1226BA0008).
\end{acknowledgments}

\section*{Data Availability}
The data that support the findings of this study are available from the corresponding author upon reasonable request.

\nocite{*}
\bibliographystyle{apsrev4-2}
\bibliography{apssamp}

@PREAMBLE{
 "\providecommand{\noopsort}[1]{}" 
 # "\providecommand{\singleletter}[1]{#1}%" 
}

@BOOK{Bire82,
   author       = {N. D. Birell and P. C. W. Davies},
   year         = 1982,
   title        = {Quantum Fields in Curved Space},
   publisher    = {Cambridge University Press}
}

@article{2020_Koizumi,
    author = {Koizumi, Hiroki and Hagihara, Michio and Kobayashi, Soki and Yanagihara, Hideto},
    title = {Interlayer exchange coupling and interface magnetic anisotropy with crossed in-plane and perpendicular magnetic anisotropies},
    journal = {AIP Advances},
    volume = {10},
    number = {1},
    pages = {015108},
    year = {2020},
    month = {01},
    issn = {2158-3226},
    doi = {10.1063/1.5129564},
    url = {https://doi.org/10.1063/1.5129564},
}

@article{1986_Grunberg,
  title = {Layered Magnetic Structures: Evidence for Antiferromagnetic Coupling of Fe Layers across Cr Interlayers},
  author = {Gr\"unberg, P. and Schreiber, R. and Pang, Y. and Brodsky, M. B. and Sowers, H.},
  journal = {Phys. Rev. Lett.},
  volume = {57},
  issue = {19},
  pages = {2442--2445},
  numpages = {0},
  year = {1986},
  month = {Nov},
  publisher = {American Physical Society},
  doi = {10.1103/PhysRevLett.57.2442},
  url = {https://link.aps.org/doi/10.1103/PhysRevLett.57.2442}
}

@article{1956_Kasuya,
    author = {Kasuya, Tadao},
    title = {A Theory of Metallic Ferro- and Antiferromagnetism on Zener's Model},
    journal = {Progress of Theoretical Physics},
    volume = {16},
    number = {1},
    pages = {45-57},
    year = {1956},
    month = {07},
    issn = {0033-068X},
    doi = {10.1143/PTP.16.45},
    url = {https://doi.org/10.1143/PTP.16.45},
}

@article{1986_Mahkrzak,
  title = {Observation of a Magnetic Antiphase Domain Structure with Long-Range Order in a Synthetic Gd-Y Superlattice},
  author = {Majkrzak, C. F. and Cable, J. W. and Kwo, J. and Hong, M. and McWhan, D. B. and Yafet, Y. and Waszczak, J. V. and Vettier, C.},
  journal = {Phys. Rev. Lett.},
  volume = {56},
  issue = {25},
  pages = {2700--2703},
  numpages = {0},
  year = {1986},
  month = {Jun},
  publisher = {American Physical Society},
  doi = {10.1103/PhysRevLett.56.2700},
  url = {https://link.aps.org/doi/10.1103/PhysRevLett.56.2700}
}

@article{1986_Salamon,
  title = {Long-range incommensurate magnetic order in a Dy-Y multilayer},
  author = {Salamon, M. B. and Sinha, Shantanu and Rhyne, J. J. and Cunningham, J. E. and Erwin, Ross W. and Borchers, Julie and Flynn, C. P.},
  journal = {Phys. Rev. Lett.},
  volume = {56},
  issue = {3},
  pages = {259--262},
  numpages = {0},
  year = {1986},
  month = {Jan},
  publisher = {American Physical Society},
  doi = {10.1103/PhysRevLett.56.259},
  url = {https://link.aps.org/doi/10.1103/PhysRevLett.56.259}
}

@article{1987_Yafet,
    author = {Yafet, Y.},
    title = {RKKY interactions across yttrium layers in Gd‐Y superlattices},
    journal = {Journal of Applied Physics},
    volume = {61},
    number = {8},
    pages = {4058-4060},
    year = {1987},
    month = {04},
    issn = {0021-8979},
    doi = {10.1063/1.338526},
    url = {https://doi.org/10.1063/1.338526},
}

@article{1991_Bruno,
  title = {Oscillatory coupling between ferromagnetic layers separated by a nonmagnetic metal spacer},
  author = {Bruno, P. and Chappert, C.},
  journal = {Phys. Rev. Lett.},
  volume = {67},
  issue = {12},
  pages = {1602--1605},
  numpages = {0},
  year = {1991},
  month = {Sep},
  publisher = {American Physical Society},
  doi = {10.1103/PhysRevLett.67.1602},
  url = {https://link.aps.org/doi/10.1103/PhysRevLett.67.1602}
}

@article{1993_Stiles,
  title = {Exchange coupling in magnetic heterostructures},
  author = {Stiles, M. D.},
  journal = {Phys. Rev. B},
  volume = {48},
  issue = {10},
  pages = {7238--7258},
  numpages = {0},
  year = {1993},
  month = {Sep},
  publisher = {American Physical Society},
  doi = {10.1103/PhysRevB.48.7238},
  url = {https://link.aps.org/doi/10.1103/PhysRevB.48.7238}
}

@article{1999_Stiles,
title = {Interlayer exchange coupling},
journal = {Journal of Magnetism and Magnetic Materials},
volume = {200},
number = {1},
pages = {322-337},
year = {1999},
issn = {0304-8853},
doi = {https://doi.org/10.1016/S0304-8853(99)00334-0},
url = {https://www.sciencedirect.com/science/article/pii/S0304885399003340},
author = {M.D. Stiles},
}

@article{2024_Surampalli,
author = {Surampalli, Akash and Bera, Anup Kumar and Chopdekar, Rajesh Vilas and Kalitsov, Alan and Wan, Lei and Katine, Jordan and Stewart, Derek and Santos, Tiffany and Prasad, Bhagwati},
title = {Voltage Controlled Interlayer Exchange Coupling and Magnetic Anisotropy Effects in Perpendicular Magnetic Heterostructures},
journal = {Advanced Functional Materials},
volume = {34},
number = {51},
pages = {2408599},
doi = {https://doi.org/10.1002/adfm.202408599},
url = {https://advanced.onlinelibrary.wiley.com/doi/abs/10.1002/adfm.202408599},
year = {2024}
}

@article{2010_Yu,
    title = {Interface Ferromagnetism and Orbital Reconstruction in {${\mathrm{BiFeO}}_{3}-{\mathrm{La}}_{0.7}{\mathrm{Sr}}_{0.3}{\mathrm{MnO}}_{3}$} Heterostructures},
    author = {Yu, P. and Lee, J.-S. and Okamoto, S. and Rossell, M. D. and Huijben, M. and Yang, C.-H. and He, Q. and Zhang, J. X. and Yang, S. Y. and Lee, M. J. and Ramasse, Q. M. and Erni, R. and Chu, Y.-H. and Arena, D. A. and Kao, C.-C. and Martin, L. W. and Ramesh, R.},
    journal = {Phys. Rev. Lett.},
    volume = {105},
    issue = {2},
    pages = {027201},
    numpages = {5},
    year = {2010},
    month = {Jul},
    publisher = {American Physical Society},
    doi = {10.1103/PhysRevLett.105.027201},
    url = {https://link.aps.org/doi/10.1103/PhysRevLett.105.027201}
}

@article{1999_Nogues,
title = {Exchange bias},
journal = {Journal of Magnetism and Magnetic Materials},
volume = {192},
number = {2},
pages = {203-232},
year = {1999},
issn = {0304-8853},
doi = {https://doi.org/10.1016/S0304-8853(98)00266-2},
url = {https://www.sciencedirect.com/science/article/pii/S0304885398002662},
author = {J Nogués and Ivan K Schuller},
}

@article{1988_Hasegawa,
  title = {Finite-temperature band theory of ferromagnetic-antiferromagnetic interfaces including exchange anisotropy},
  author = {Hasegawa, Hideo and Herman, Frank},
  journal = {Phys. Rev. B},
  volume = {38},
  issue = {7},
  pages = {4863--4872},
  numpages = {0},
  year = {1988},
  month = {Sep},
  publisher = {American Physical Society},
  doi = {10.1103/PhysRevB.38.4863},
  url = {https://link.aps.org/doi/10.1103/PhysRevB.38.4863}
}

@article{1995_Bruno,
  title = {Theory of interlayer magnetic coupling},
  author = {Bruno, P.},
  journal = {Phys. Rev. B},
  volume = {52},
  issue = {1},
  pages = {411--439},
  numpages = {0},
  year = {1995},
  month = {Jul},
  publisher = {American Physical Society},
  doi = {10.1103/PhysRevB.52.411},
  url = {https://link.aps.org/doi/10.1103/PhysRevB.52.411}
}

@article{1970_Brodkorb,
author = {Brodkorb, W. and Haubenreisser, W.},
title = {Switching curves of exchange-coupled ferro—antiferromagnetic double layer films},
journal = {physica status solidi (a)},
volume = {3},
number = {2},
pages = {333-341},
doi = {https://doi.org/10.1002/pssa.19700030207},
url = {https://onlinelibrary.wiley.com/doi/abs/10.1002/pssa.19700030207},
year = {1970}
}

@article{1997_Koon,
  title = {Calculations of Exchange Bias in Thin Films with Ferromagnetic/Antiferromagnetic Interfaces},
  author = {Koon, N. C.},
  journal = {Phys. Rev. Lett.},
  volume = {78},
  issue = {25},
  pages = {4865--4868},
  numpages = {0},
  year = {1997},
  month = {Jun},
  publisher = {American Physical Society},
  doi = {10.1103/PhysRevLett.78.4865},
  url = {https://link.aps.org/doi/10.1103/PhysRevLett.78.4865}
}

@article{1994_Dieny,
title = {Giant magnetoresistance in spin-valve multilayers},
journal = {Journal of Magnetism and Magnetic Materials},
volume = {136},
number = {3},
pages = {335-359},
year = {1994},
issn = {0304-8853},
doi = {https://doi.org/10.1016/0304-8853(94)00356-4},
author = {B. Dieny},
}

@article{1996_Moran,
    author = {Moran, Timothy J. and Schuller, Ivan K.},
    title = {Effects of cooling field strength on exchange anisotropy at permalloy/CoO interfaces},
    journal = {Journal of Applied Physics},
    volume = {79},
    number = {8},
    pages = {5109-5111},
    year = {1996},
    month = {04},
    issn = {0021-8979},
    doi = {10.1063/1.361317},
    url = {https://doi.org/10.1063/1.361317},
}

@article{1992_Johnson,
  title = {Structural dependence of the oscillatory exchange interaction across Cu layers},
  author = {Johnson, M. T. and Purcell, S. T. and McGee, N. W. E. and Coehoorn, R. and aan de Stegge, J. and Hoving, W.},
  journal = {Phys. Rev. Lett.},
  volume = {68},
  issue = {17},
  pages = {2688--2691},
  numpages = {0},
  year = {1992},
  month = {Apr},
  publisher = {American Physical Society},
  doi = {10.1103/PhysRevLett.68.2688},
  url = {https://link.aps.org/doi/10.1103/PhysRevLett.68.2688}
}

@article{2006_Katayama,
    author = {Katayama, T. and Yuasa, S. and Velev, J. and Zhuravlev, M. Ye. and Jaswal, S. S. and Tsymbal, E. Y.},
    title = {Interlayer exchange coupling in FeMgOFe magnetic tunnel junctions},
    journal = {Applied Physics Letters},
    volume = {89},
    number = {11},
    pages = {112503},
    year = {2006},
    month = {09},
    issn = {0003-6951},
    doi = {10.1063/1.2349321},
    url = {https://doi.org/10.1063/1.2349321},
}

@article{2015_Yanagihara,
  title = {Antiferromagnetic coupling and impurity effects at junctions between ${\mathbf{Fe}}_{\mathbf{3}}{\mathbf{O}}_{\mathbf{4}}$ and Fe(001) layers},
  author = {Yanagihara, H. and Kamita, H. and Honda, S. and Inoue, J. and Kita, Eiji and Itoh, H. and Mibu, Ko},
  journal = {Phys. Rev. B},
  volume = {91},
  issue = {17},
  pages = {174423},
  numpages = {6},
  year = {2015},
  month = {May},
  publisher = {American Physical Society},
  doi = {10.1103/PhysRevB.91.174423},
  url = {https://link.aps.org/doi/10.1103/PhysRevB.91.174423}
}

@article{2012_Bailey,
  title = {Pd magnetism induced by indirect interlayer exchange coupling},
  author = {Bailey, W. E. and Ghosh, A. and Auffret, S. and Gautier, E. and Ebels, U. and Wilhelm, F. and Rogalev, A.},
  journal = {Phys. Rev. B},
  volume = {86},
  issue = {14},
  pages = {144403},
  numpages = {5},
  year = {2012},
  month = {Oct},
  publisher = {American Physical Society},
  doi = {10.1103/PhysRevB.86.144403},
  url = {https://link.aps.org/doi/10.1103/PhysRevB.86.144403}
}

@article{1990_Parkin,
  title = {Oscillations in exchange coupling and magnetoresistance in metallic superlattice structures: Co/Ru, Co/Cr, and Fe/Cr},
  author = {Parkin, S. S. P. and More, N. and Roche, K. P.},
  journal = {Phys. Rev. Lett.},
  volume = {64},
  issue = {19},
  pages = {2304--2307},
  numpages = {0},
  year = {1990},
  month = {May},
  publisher = {American Physical Society},
  doi = {10.1103/PhysRevLett.64.2304},
  url = {https://link.aps.org/doi/10.1103/PhysRevLett.64.2304}
}

@article{2025_Wang,
  title = {Dynamical synthetic antiferromagnetic skyrmions and skyrmionia},
  author = {Wang, Shun and Yao, Linrong and Zhou, Yan and Jiang, Sheng},
  journal = {Phys. Rev. Appl.},
  volume = {24},
  issue = {3},
  pages = {034054},
  numpages = {14},
  year = {2025},
  month = {Sep},
  publisher = {American Physical Society},
  doi = {10.1103/jssf-k5yr},
  url = {https://link.aps.org/doi/10.1103/jssf-k5yr}
}

@article{2013_Niizeki,
    author = {Niizeki, Tomohiko and Utsumi, Yuji and Aoyama, Ryohei and Yanagihara, Hideto and Inoue, Jun-ichiro and Yamasaki, Yuichi and Nakao, Hironori and Koike, Kazuyuki and Kita, Eiji},
    title = {Extraordinarily large perpendicular magnetic anisotropy in epitaxially strained cobalt ferrite {Co$_{x}$Fe$_{3-x}$O$_4$(001)} (x=0.75, 1.0) thin films},
    journal = {Applied Physics Letters},
    volume = {103},
    number = {16},
    pages = {162407},
    year = {2013},
    month = {10},
    issn = {0003-6951},
    doi = {10.1063/1.4824761},
    url = {https://doi.org/10.1063/1.4824761},
}

@article{2022_RUIZGOMEZ,
title = {Magnetic domain wall pinning in cobalt ferrite microstructures},
journal = {Applied Surface Science},
volume = {600},
pages = {154045},
year = {2022},
issn = {0169-4332},
doi = {https://doi.org/10.1016/j.apsusc.2022.154045},
url = {https://www.sciencedirect.com/science/article/pii/S0169433222015835},
author = {Sandra Ruiz-Gómez and Anna Mandziak and Laura Martín-García and José Emilio Prieto and Pilar Prieto and Carmen Munuera and Michael Foerster and Adrián Quesada and Lucía Aballe and Juan {de la Figuera}},
}

@incollection{1995_BRABERS,
title = {Chapter 3 Progress in spinel ferrite research},
booktitle = {Handbook of Magnetic Materials},
publisher = {Elsevier},
volume = {8},
pages = {189-324},
year = {1995},
issn = {1567-2719},
doi = {https://doi.org/10.1016/S1567-2719(05)80032-0},
url = {https://www.sciencedirect.com/science/article/pii/S1567271905800320},
author = {V.A.M. Brabers},
}

@article{2017_Coey,
  title = {Magnetization and anisotropy of cobalt ferrite thin films},
  author = {Eskandari, F. and Porter, S. B. and Venkatesan, M. and Kameli, P. and Rode, K. and Coey, J. M. D.},
  journal = {Phys. Rev. Mater.},
  volume = {1},
  issue = {7},
  pages = {074413},
  numpages = {10},
  year = {2017},
  month = {Dec},
  publisher = {American Physical Society},
  doi = {10.1103/PhysRevMaterials.1.074413},
  url = {https://link.aps.org/doi/10.1103/PhysRevMaterials.1.074413}
}

@article{1990_Cain,
    author = {Cain, William C. and Kryder, Mark H.},
    title = {Investigation of the exchange mechanism in NiFe‐TbCo bilayers},
    journal = {Journal of Applied Physics},
    volume = {67},
    number = {9},
    pages = {5722-5724},
    year = {1990},
    month = {05},
    issn = {0021-8979},
    doi = {10.1063/1.346107},
    url = {https://doi.org/10.1063/1.346107},
    eprint = {https://pubs.aip.org/aip/jap/article-pdf/67/9/5722/18634176/5722\_1\_online.pdf},
}

@article{2005_Katsura,
  title = {Spin Current and Magnetoelectric Effect in Noncollinear Magnets},
  author = {Katsura, Hosho and Nagaosa, Naoto and Balatsky, Alexander V.},
  journal = {Phys. Rev. Lett.},
  volume = {95},
  issue = {5},
  pages = {057205},
  numpages = {4},
  year = {2005},
  month = {Jul},
  publisher = {American Physical Society},
  doi = {10.1103/PhysRevLett.95.057205},
  url = {https://link.aps.org/doi/10.1103/PhysRevLett.95.057205}
}

@article{2023_Liu,
title = {Noncollinear interlayer exchange coupling across IrFe spacer layers},
journal = {Journal of Magnetism and Magnetic Materials},
volume = {585},
pages = {171109},
year = {2023},
issn = {0304-8853},
doi = {https://doi.org/10.1016/j.jmmm.2023.171109},
url = {https://www.sciencedirect.com/science/article/pii/S030488532300759X},
author = {Juliana Lisik and Spencer Myrtle and Erol Girt}
}

@article{2004_Stobiecki,
author = {Stobiecki, T. and Kanak, J. and Wrona, J. and Czapkiewicz, M. and Kim, C. G. and Kim, C. O. and Tsunoda, M. and Takahashi, M.},
title = {Correlation between structure and exchange coupling parameters of IrMn based MTJ},
journal = {physica status solidi (a)},
volume = {201},
number = {8},
pages = {1621-1627},
year = {2004},
doi = {https://doi.org/10.1002/pssa.200304661},
}

@article{2014_Nguyen,
  title = {Depth-Dependent Magnetization Profiles of Hybrid Exchange Springs},
  author = {Nguyen, T. N. Anh and Knut, R. and Fallahi, V. and Chung, S. and Le, Q. Tuan and Mohseni, S. M. and Karis, O. and Peredkov, S. and Dumas, R. K. and Miller, Casey W. and \AA{}kerman, J.},
  journal = {Phys. Rev. Appl.},
  volume = {2},
  issue = {4},
  pages = {044014},
  numpages = {7},
  year = {2014},
  month = {Oct},
  publisher = {American Physical Society},
  doi = {10.1103/PhysRevApplied.2.044014},
  url = {https://link.aps.org/doi/10.1103/PhysRevApplied.2.044014}
}

@article{2014_Vansteenkiste_Mumax3,
    author  = {Vansteenkiste, Arne and
               Leliaert, Jonathan and
               Dvornik, Mykola and
               Helsen, Mathias and
               Garcia-Sanchez, Felipe and
               {Van Waeyenberge}, Bartel},
    title   = {{The design and verification of Mumax3}},
    journal = {AIP Advances},
    number  = {10},
    pages   = {107133},
    volume  = {4},
    year    = {2014},
    doi     = {10.1063/1.4899186},
    url     = {http://doi.org/10.1063/1.4899186}
}

@article{2011_Nguyen,
    author = {Nguyen, T. N. Anh and Fang, Y. and Fallahi, V. and Benatmane, N. and Mohseni, S. M. and Dumas, R. K. and Åkerman, Johan},
    title = {[Co/Pd]–NiFe exchange springs with tunable magnetization tilt angle},
    journal = {Applied Physics Letters},
    volume = {98},
    number = {17},
    pages = {172502},
    year = {2011},
    month = {04},
    issn = {0003-6951},
    doi = {10.1063/1.3580612},
    url = {https://doi.org/10.1063/1.3580612},
}

@article{1989_Grunberg,
  title = {Enhanced magnetoresistance in layered magnetic structures with antiferromagnetic interlayer exchange},
  author = {Binasch, G. and Gr\"unberg, P. and Saurenbach, F. and Zinn, W.},
  journal = {Phys. Rev. B},
  volume = {39},
  issue = {7},
  pages = {4828--4830},
  numpages = {0},
  year = {1989},
  month = {Mar},
  publisher = {American Physical Society},
  doi = {10.1103/PhysRevB.39.4828},
  url = {https://link.aps.org/doi/10.1103/PhysRevB.39.4828}
}

@article{2020_olmos,
    author = {Olmos, Rubyann and Iturriaga, Hector and Blazer, Dawn S. and Koohfar, Sanaz and Gandha, Kinjal and Nlebedim, Ikenna C. and Kumah, Divine P. and Singamaneni, Srinivasa R.},
    title = {Exchange bias in La0.7Sr0.3CrO3/La0.7Sr0.3MnO3/La0.7Sr0.3CrO3 heterostructures},
    journal = {AIP Advances},
    volume = {10},
    number = {1},
    pages = {015001},
    year = {2020},
    month = {01},
    issn = {2158-3226},
    doi = {10.1063/1.5130453},
}

@article{2018_kuswik,
  title = {Asymmetric domain wall propagation caused by interfacial Dzyaloshinskii-Moriya interaction in exchange biased Au/Co/NiO layered system},
  author = {Ku\ifmmode \acute{s}\else \'{s}\fi{}wik, P. and Matczak, M. and Kowacz, M. and Szuba-Jab\l{}o\ifmmode \acute{n}\else \'{n}\fi{}ski, K. and Michalak, N. and Szyma\ifmmode \acute{n}\else \'{n}\fi{}ski, B. and Ehresmann, A. and Stobiecki, F.},
  journal = {Phys. Rev. B},
  volume = {97},
  issue = {2},
  pages = {024404},
  numpages = {7},
  year = {2018},
  month = {Jan},
  publisher = {American Physical Society},
  doi = {10.1103/PhysRevB.97.024404},
  url = {https://link.aps.org/doi/10.1103/PhysRevB.97.024404}
}

@article{2017_Belmeguenai,
  title = {Interface Dzyaloshinskii-Moriya interaction in the interlayer antiferromagnetic-exchange coupled Pt/CoFeB/Ru/CoFeB systems},
  author = {Belmeguenai, M. and Bouloussa, H. and Roussign\'e, Y. and Gabor, M. S. and Petrisor, T. and Tiusan, C. and Yang, H. and Stashkevich, A. and Ch\'erif, S. M.},
  journal = {Phys. Rev. B},
  volume = {96},
  issue = {14},
  pages = {144402},
  numpages = {7},
  year = {2017},
  month = {Oct},
  publisher = {American Physical Society},
  doi = {10.1103/PhysRevB.96.144402},
  url = {https://link.aps.org/doi/10.1103/PhysRevB.96.144402}
}

@ARTICLE{2014_Yoshida,
  author={Yoshida, Chikako and Takenaga, Takashi and Yamazaki, Yuichi and Uehara, Haruka and Noshiro, Hideyuki and Tsunoda, Koji and Iba, Yoshihisa and Hatada, Akiyoshi and Nakabayashi, Masaaki and Takahashi, Atsushi and Aoki, Masaki and Sugii, Toshihiro},
  journal={IEEE Transactions on Magnetics}, 
  title={Reduction of Offset Field in Top-Pinned MTJ With Synthetic Antiferromagnetic Free Layer}, 
  year={2014},
  volume={50},
  number={11},
  pages={1-4},
  doi={10.1109/TMAG.2014.2325965}}

@article{2024_Nakatani,
    author = {Kulkarni, Prabhanjan D. and Nakatani, Tomoya},
    title = {Tunnel magnetoresistive sensors with non-hysteretic resistance–magnetic field curves using noncollinear interlayer exchange coupling through RuFe spacers},
    journal = {Applied Physics Letters},
    volume = {125},
    number = {16},
    pages = {162405},
    year = {2024},
    month = {10},
    issn = {0003-6951},
    doi = {10.1063/5.0231451},
    url = {https://doi.org/10.1063/5.0231451},
}

@article{2002_Liu,
    author = {Liu, Xiaoyong and Ren, Cong and Xiao, Gang},
    title = {Magnetic tunnel junction field sensors with hard-axis bias field},
    journal = {Journal of Applied Physics},
    volume = {92},
    number = {8},
    pages = {4722-4725},
    year = {2002},
    month = {10},
    issn = {0021-8979},
    doi = {10.1063/1.1507818},
    url = {https://doi.org/10.1063/1.1507818},
}

@article{2020_Zachary,
author = {Zachary R. Nunn  and Claas Abert  and Dieter Suess  and Erol Girt },
title = {Control of the noncollinear interlayer exchange coupling},
journal = {Science Advances},
volume = {6},
number = {48},
pages = {eabd8861},
year = {2020},
doi = {10.1126/sciadv.abd8861},
URL = {https://www.science.org/doi/abs/10.1126/sciadv.abd8861},
}

@article{2023_Lisik,
title = {Noncollinear interlayer exchange coupling across IrFe spacer layers},
journal = {Journal of Magnetism and Magnetic Materials},
volume = {585},
pages = {171109},
year = {2023},
issn = {0304-8853},
doi = {https://doi.org/10.1016/j.jmmm.2023.171109},
url = {https://www.sciencedirect.com/science/article/pii/S030488532300759X},
author = {Juliana Lisik and Spencer Myrtle and Erol Girt},
}

@article{1990_Dieny,
doi = {10.1088/0953-8984/2/1/013},
url = {https://doi.org/10.1088/0953-8984/2/1/013},
year = {1990},
month = {jan},
publisher = {},
volume = {2},
number = {1},
pages = {159},
author = {B Dieny and J P Gavigan and J P Rebouillat},
title = {Magnetisation processes, hysteresis and finite-size effects in model multilayer systems of cubic or uniaxial anisotropy with antiferromagnetic coupling between adjacent ferromagnetic layers},
journal = {Journal of Physics: Condensed Matter},
}

@article{2015_Lee,
    author = {Lee, Y. C. and Chao, C. T. and Li, L. C. and Suen, Y. W. and Horng, Lance and Wu, Te-Ho and Chang, C. R. and Wu, J. C.},
    title = {Magnetic tunnel junction based out-of-plane field sensor with perpendicular magnetic anisotropy in reference layer},
    journal = {Journal of Applied Physics},
    volume = {117},
    number = {17},
    pages = {17A320},
    year = {2015},
    month = {03},
    issn = {0021-8979},
    doi = {10.1063/1.4914121},
    url = {https://doi.org/10.1063/1.4914121},
}

@article{1990_Cahill,
    author = {Cahill, David G.},
    title = {Thermal conductivity measurement from 30 to 750 K: the 3ω method},
    journal = {Review of Scientific Instruments},
    volume = {61},
    number = {2},
    pages = {802-808},
    year = {1990},
    month = {02},
    issn = {0034-6748},
    doi = {10.1063/1.1141498},
    url = {https://doi.org/10.1063/1.1141498},
}

@article{2017_Kimling,
  title = {Thermal conductance of interfaces with amorphous ${\mathbf{SiO}}_{2}$ measured by time-resolved magneto-optic Kerr-effect thermometry},
  author = {Kimling, Judith and Philippi-Kobs, Andr\'e and Jacobsohn, Jonathan and Oepen, Hans Peter and Cahill, David G.},
  journal = {Phys. Rev. B},
  volume = {95},
  issue = {18},
  pages = {184305},
  numpages = {10},
  year = {2017},
  month = {May},
  publisher = {American Physical Society},
  doi = {10.1103/PhysRevB.95.184305},
  url = {https://link.aps.org/doi/10.1103/PhysRevB.95.184305}
}

@article{2025_He,
    author = {He, Rongjie and Sun, Bo},
    title = {Thermal conductivity of SiO2 grown by plasma enhanced chemical vapor deposition},
    journal = {Journal of Applied Physics},
    volume = {137},
    number = {17},
    pages = {175109},
    year = {2025},
    month = {05},
    issn = {0021-8979},
    doi = {10.1063/5.0246816},
    url = {https://doi.org/10.1063/5.0246816},
}

@article{1972_Ho,
    author = {Ho, C. Y. and Powell, R. W. and Liley, P. E.},
    title = {Thermal Conductivity of the Elements},
    journal = {Journal of Physical and Chemical Reference Data},
    volume = {1},
    number = {2},
    pages = {279-421},
    year = {1972},
    month = {04},
    issn = {0047-2689},
    doi = {10.1063/1.3253100},
    url = {https://doi.org/10.1063/1.3253100},
}

@article{2018_Zhang,
    author = {Zhang, Chen and Huberman, Samuel C. and Ning, Shuai and Pelliciari, Jonathan and Duncan, Ryan A. and Liao, Bolin and Ojha, Shuchi and Freeland, John W. and Nelson, Keith A. and Comin, Riccardo and Chen, Gang and Ross, Caroline A.},
    title = {Thermal conductivity in self-assembled CoFe2O4/BiFeO3 vertical nanocomposite films},
    journal = {Applied Physics Letters},
    volume = {113},
    number = {22},
    pages = {223105},
    year = {2018},
    month = {11},
    issn = {0003-6951},
    doi = {10.1063/1.5049176},
    url = {https://doi.org/10.1063/1.5049176},
}

@article{2016_Jiang,
    author = {Jiang, Puqing and Huang, Bin and Koh, Yee Kan},
    title = {Accurate measurements of cross-plane thermal conductivity of thin films by dual-frequency time-domain thermoreflectance (TDTR)},
    journal = {Review of Scientific Instruments},
    volume = {87},
    number = {7},
    pages = {075101},
    year = {2016},
    month = {07},
    issn = {0034-6748},
    doi = {10.1063/1.4954969},
    url = {https://doi.org/10.1063/1.4954969},
}

\clearpage
\onecolumngrid
\linespread{1.6}\selectfont

\begin{center}
\textbf{\large Supplement materials to "Probing Spin Configurations in Exchange-Coupled Magnetic Bilayers with Orthogonal Anisotropies via Anomalous Hall and Nernst Effects"}
\end{center}
\vspace{4ex}

\setcounter{section}{0}
\setcounter{figure}{0}
\setcounter{table}{0}
\setcounter{equation}{0}
\renewcommand{\thefigure}{S\arabic{figure}}
\renewcommand{\thetable}{S\arabic{table}}
\renewcommand{\theequation}{S\arabic{equation}}

\section{Structural Characterization}

\begin{figure}[htbp]
    \centering
    \includegraphics[width=0.6\columnwidth]{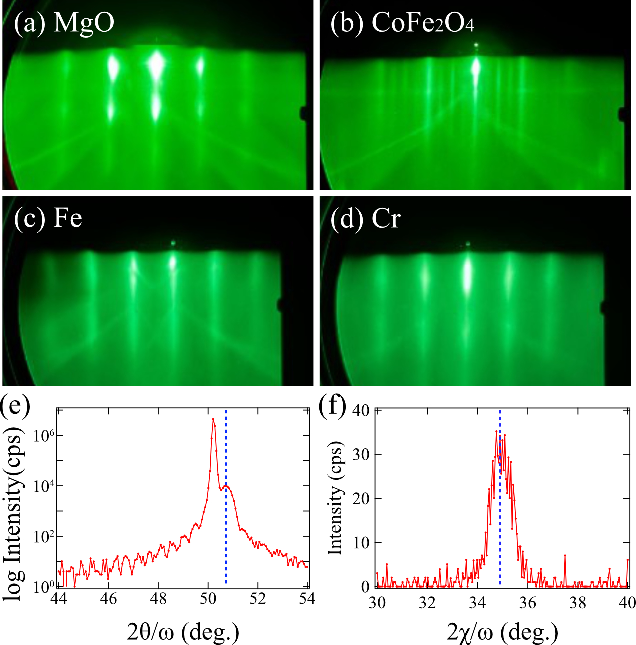} 
    \caption{Crystallinity of the grown stack of sample measured with RHEED and X-ray diffraction (XRD) spectroscopy. (a-d) In-situ RHEED patterns for each surface. (e, f) Out-of-plane and in-plane XRD patterns, respectively.}
    \label{figs1}
\end{figure}

Figure \ref{figs1} presents the structural characterization of the samples using in situ RHEED and X-ray diffraction (XRD). The RHEED patterns taken after the growth of each layer [Figs. \ref{figs1}(a–d)] confirmed the epitaxial growth of both the CoFe$_2$O$_4$ and Fe layers on the MgO substrate. The observation of distinct Kikuchi lines indicated a high degree of surface flatness for both layers. Notably, the RHEED pattern for the CoFe$_2$O$_4$ surface exhibits additional streaks characteristic of its spinel structure.

XRD measurements were performed to evaluate the crystallinity of the bilayers(Fig. \ref{figs1}(e, f)). The blue dashed line in out-of-plane XRD profile (Fig. \ref{figs1}(e)) showed the CoFe$_2$O$_4$ (004) peak. Also, the distinct fringes around the CoFe$_2$O$_4$ (004) reflection was observed, indicating a high degree of surface flatness for the CoFe$_2$O$_4$ layer. In the in-plane XRD (Fig. \ref{figs1}(f)), the CoFe$_2$O$_4$ (220) peak was observed at approximately 35°, a position where the MgO (110) reflection is forbidden by structural selection rules. This observation suggested that the CoFe$_2$O$_4$ layer crystallized in a spinel structure. Overall, the RHEED and XRD result demonstrated the high crystalline quality of the CoFe$_2$O$_4$/Fe bilayers.

\section{ANE measurements for y-direction magnetization}
\begin{figure}[htbp]
    \centering
    \includegraphics[width=0.6\columnwidth]{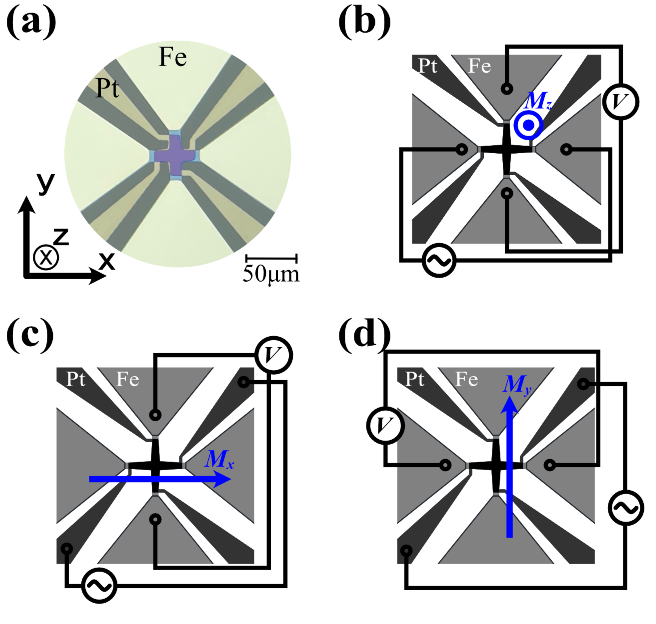} 
    \caption{Device pattern and measurement configurations for combined AHE and ANE measurements. (a) Optical micrograph of the patterned device. The dimensions of the Hall bar and the Pt heater are 10 × 50 $\rm{\mu}$m² and 10 × 40 $\rm{\mu}$m², respectively. The both layers are electrically separated by SiO$_2$ layers. (b) Measurement configuration for the AHE. (c, d) Measurement configurations for the ANE. By modifying the current path and voltage contacts, the in-plane components of the magnetization ($M_x$ and $M_y$) can be independently probed.}
    \label{figs2}
\end{figure}

Using a Hall bar device with an integrated Pt heater, the three-dimensional magnetization components of the conductive layer can be accessed by simply switching the current path and voltage contacts. The dimensions of the Hall bar and the Pt heater are 10 × 50 $\mu$m² and 10 × 40 $\mu$m², respectively (Fig. \ref{figs2}(a)). By applying a current directly to the Fe layer and measuring the transverse voltage, the AHE signal was obtained (Fig. \ref{figs2}(b)). When the current path was switched to the Pt heater, Joule heating generated a temperature gradient along the z-direction of the Hall bar. This allowed the in-plane magnetization components ($M_x$ and $M_y$) to be independently probed via the ANE (Figs. \ref{figs2}(c, d)).

\begin{figure}[htbp]
    \centering
    \includegraphics[width=1.0\columnwidth]{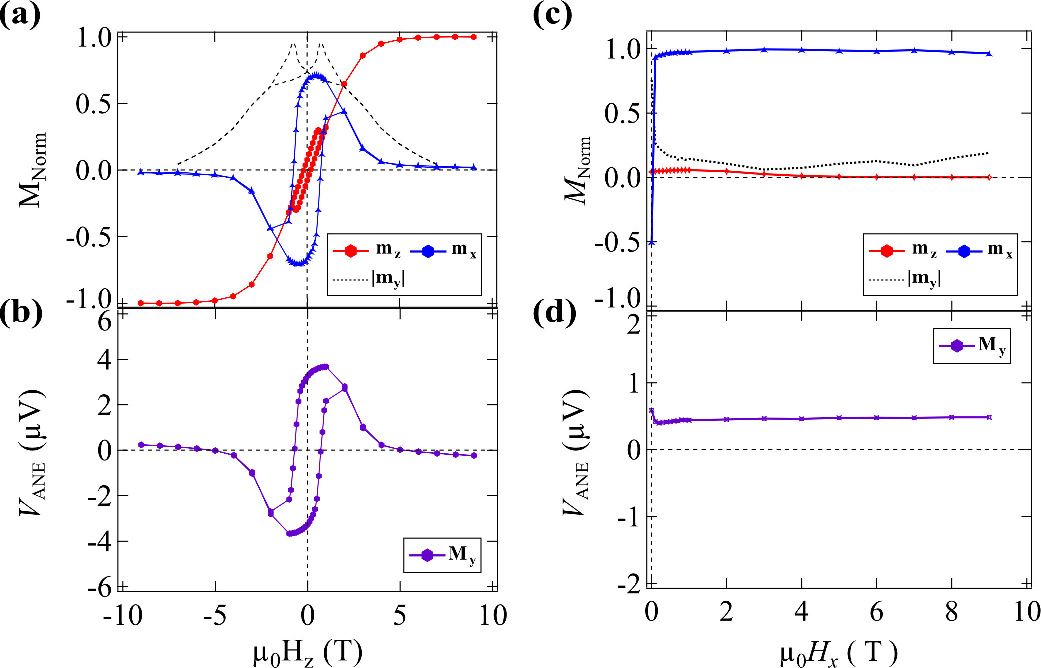} 
    \caption{The measured three dimensional magnetization information under out-of-plane and in-plane magnetic field. (a) The normalized $M_x$, $M_z$ components measured using AHE and ANE measurements under in-plane magnetic field. Under the unit vector assumption, the absolute value of $M_y$ component was calculated. (b) The measured $M_y$ component using ANE measurement configuration shown in Fig. \ref{figs2} (d) under out-of-plane magnetic field. (c) The normalized $M_x$, $M_z$ components measured using AHE and ANE measurements under in-plane magnetic field, with calculated $|M_y|$ under unit vector assumption. (d) The measured $M_y$ component using ANE measurement configuration shown in Fig. \ref{figs2} (d) under in-plane magnetic field.}
    \label{figs3}
\end{figure}

Figures \ref{figs3}(a) and (c) present the normalized magnetization components ($M_x$ and $M_z$) obtained from combined AHE and ANE measurements and the $|M_y|$ component calculated using a unit-vector assumption under out-of-plane and in-plane magnetic fields, respectively. Figures \ref{figs3} (b) and (d) present the ANE signals which were proportional to the $M_y$ with measurement configuration shown in Fig. \ref{figs2}(d). Although the lack of an absolute magnitude for saturated $M_y$ preclude its direct normalization, the ANE measurements reliably captured the signal profile. The close agreement between the calculated $|M_y|$ profile and the measured ANE signal validates the macrospin model, confirming that the $M_y$ component can be evaluated using this methodology.

\section{Effect of interlayer exchange coupling on the hump-shaped AHE signal}
\begin{figure}[htbp]
    \centering
    \includegraphics[width=0.8\columnwidth]{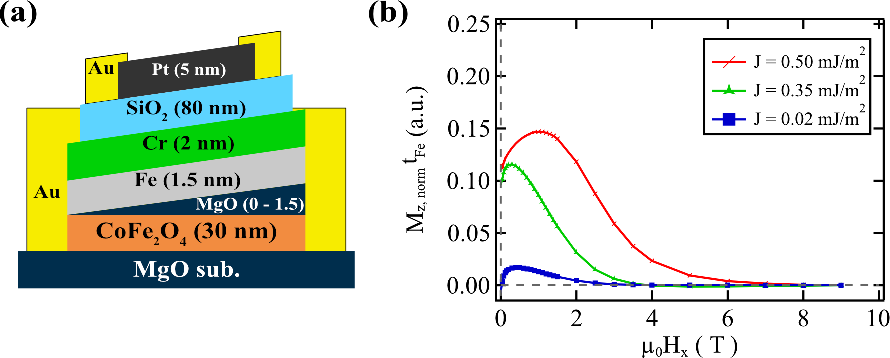} 
    \caption{Modulation of the hump-shaped AHE signal by interlayer exchange coupling. (a) Schematic of the sample structure with an MgO spacer inserted between the CoFe$_2$O$_4$ and Fe layers. (b) AHE curves measured for varying MgO spacer thicknesses under in-plane magnetic field (0, 0.5, and 1.0 nm). The intensity of the hump-shaped signal decreases as the spacer thickness increases, demonstrating that the phenomenon was driven by interfacial antiferromagnetic exchange coupling.}
    \label{figs5}
\end{figure}

To further investigate the effect of exchange coupling on the hump-shaped AHE signal, we inserted an MgO spacer between the CoFe$_2$O$_4$ and Fe layers to weaken the interlayer exchange coupling\cite{2020_Koizumi}. For this measurement, the rest of the structure, excluding the MgO spacer, was fabricated and lithographically patterned under the same conditions as previously described, as shown in Fig. \ref{figs5}(a). From the AHE measurements under an out-of-plane magnetic field, the exchange coupling constants were determined by equating the Zeeman energy and the exchange coupling energy between the remanent state and the coercive field.

Using these samples, the effect of the exchange coupling constant on the hump-shaped AHE signal under an in-plane magnetic field was measured (Fig. \ref{figs5}(b)). For an MgO spacer thickness of 0 nm, the exchange coupling constant matched the previous results, and exhibited the same hump-shaped signal. When the thickness of the MgO spacer was 0.5 nm, the exchange coupling constant decreased to 0.35 mJ/m$^2$, accompanied by a reduction in the intensity of the AHE signal under the in-plane magnetic field. When the spacer thickness exceeded 1.0 nm, the exchange coupling vanished, resulting in the total disappearance of the hump-shaped signal. This result further supports that the phenomenon stems from the antiferromagnetic exchange coupling at the interface.

\section{Evaluation of the induced thermal gradient}
In the Hall bar device with an integrated Pt heater, applying an electrical current generated a $z$-axis heat flux across the CoFe$_2$O$_4$/Fe bilayer. Since the ANE signals originate exclusively from the Fe layer, the induced temperature gradient can be determined using the ANE coefficient of Fe. When a current of 20 mA was applied to the Pt layer, an ANE signal of 2.3 µV was measured (Fig. 3(b)). By accounting for ANE coefficient of Fe and the device structure, we evaluated the temperature difference generated across the Fe layer to be 0.7 mK.

\begin{figure}[htbp]
    \centering
    \includegraphics[width=0.6\columnwidth]{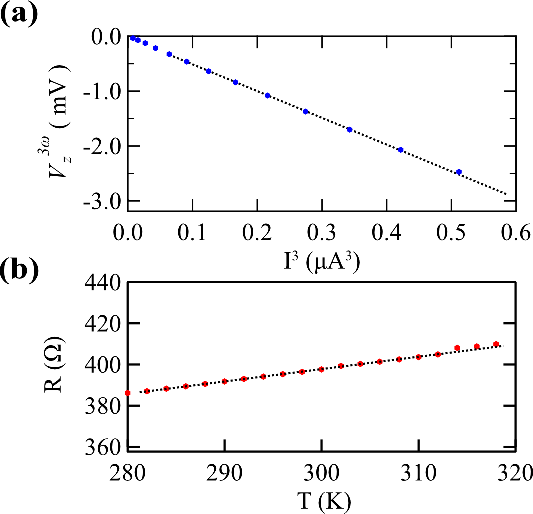} 
    \caption{Current dependence of 3$\omega$ component and temperature dependence of sample resistivity for determining the temperature increase using 3$\omega$ method. (a) Current dependence of 3$\omega$ component for vertical resistivity. (b) The temperature dependence of sample resistivity.}
    \label{figs4}
\end{figure}

The temperature difference generated across the Fe layer can also be roughly estimated using Fourier's law of heat conduction. The temperature increase in the Pt heater was determined via the $3\omega$ method\cite{1990_Cahill} as shown in Fig. \ref{figs4}. From the current dependence of 3$\omega$ component(Fig. \ref{figs4}(a)) and temperature dependency of the resistivity of Pt layer(Fig. \ref{figs4}(b)), $\alpha = 0.6 \ \Omega\cdot\text{K}^{-1}$ and $\beta = 30.5 \ \text{mK}/I^2$ could be determined where

\begin{equation}
\begin{split}
R(T) &=R_0 + \alpha T,\ T(I) = T_0 + \beta I^2\\
V_Z^{3\omega} &=-\frac{1}{4}\alpha \beta I_0^3\sin(3\omega t).
\end{split}
\label{equs1}
\end{equation}

Applying a current of 20 mA to the Pt layer induced a temperature increase of 12 K. Assuming the substrate acted as a heat bath, the temperature difference across the Fe layer was calculated to be 3.5 mK based on Fourier's law with parameters on table \ref{tableS1}. Both rough assumption yielded closely matching results, suggesting that a reasonable ANE signal was measured from the ANE measurement.

\begin{table}[htbp]
\centering
\begin{tabular}{lccc}
\toprule
 & \textbf{Thermal resistivity (m$\cdot$K/W)} & \textbf{Thermal resistance ($10^{-9}$$\rm m^2$W/K)} & \textbf{Reference} \\
\midrule
$SiO_2$ & 0.68 & 54 & \cite{2025_He} \\
$Fe$ & 0.013 & 0.020 & \cite{1972_Ho} \\
$CoFe_2O_4$ & 0.25 & 7.5 & \cite{2018_Zhang} \\
$SiO_2/Pt$ interface& . & 3.3 & \cite{2017_Kimling} \\
$SiO_2/Fe$ interface & . & 3.7 & \cite{2016_Jiang} \\
\bottomrule
\end{tabular}
\caption{Thermal parameter for calculating the Fourier's law of thermal resistivity. Thermal resistance was calculated with device structure and thermal resistivity. The effect of the interface were also considered to calculate the thermal properties.}
\label{tableS1}
\end{table}

\end{document}